\documentclass[12pt]{iopart}

\usepackage{graphicx} 
\usepackage{dcolumn} 
\usepackage{bm} 

\newcommand{ \be }{\begin{equation}}    
\newcommand{ \ee }{\end{equation}}    
\newcommand{ \bea }{\begin{eqnarray}}    
\newcommand{ \eea }{\end{eqnarray}}

\usepackage{color}

\begin{document}       

\begin{flushright}    
\end{flushright}

\title{Strangelet search at RHIC}


\author{A.H. Tang\dag\ for the STAR Collaboration\footnote{For the full author list and acknowledgements see Appendix "Collaborations" in this volume.}}
\address{\dag\ NIKHEF and BNL \\ Physics Department, P.O. Box 5000, Brookhaven National Laboratory, 
Upton, NY~11973, aihong@bnl.gov}


\date{\today}
\vspace{-0.3cm}
\begin{abstract}
Two position sensitive Shower Maximum Detector (SMDs) for Zero-Degree 
Calorimeters (ZDCs) were installed by STAR before run 2004 at both upstream 
and downstream from the interaction point along the beam axis where particles 
with small rigidity are swept away by strong magnetic field. The ZDC-SMDs provides information about neutral energy deposition as a function of transverse position in ZDCs. We report the preliminary results of strangelet search from a 
triggered data-set sampling 100 million Au+Au collisions at top RHIC energy.
\end{abstract}
\vspace{-0.2cm}


Strange Quark Matter (SQM) is a hypothetical state that consists of 
approximately equal numbers of $u$, $d$ and $s$ quarks. It has lower energy
than ordinary nuclear matter that we are familiar with, thus it might be
the true ground state of baryonic matter~\cite{Bodmer, Witten} and absolutely stable.
The search for small lumps of SQM (``strangelets'') has been performed on 
earth and in cosmos~\cite{Klingenberg}, as well in heavy ion 
experiments~\cite{E864}. So far its existence is not confirmed.

In heavy ion experiments, there are three models of strangelet production 
mechanism, namely, coalescence~\cite{Boltz},  statistical thermal production~\cite{Braun-Munzinger} and 
distillation~\cite{Greiner} via the created Quark Gluon Plasma (QGP).  
The first two models usually yield lower strangelet cross section than the 
last one. In the 
distillation process, if a QGP is created in the hot and dense fireball of
heavy ion collisions, it could lose energy by meson-emission (distillation)
and end up with a matter in its ground state -- strangelet. 

\begin{figure}[ht]
\vspace{-0.5cm}
\begin{center}
\resizebox{
\textwidth}{!}{
\includegraphics{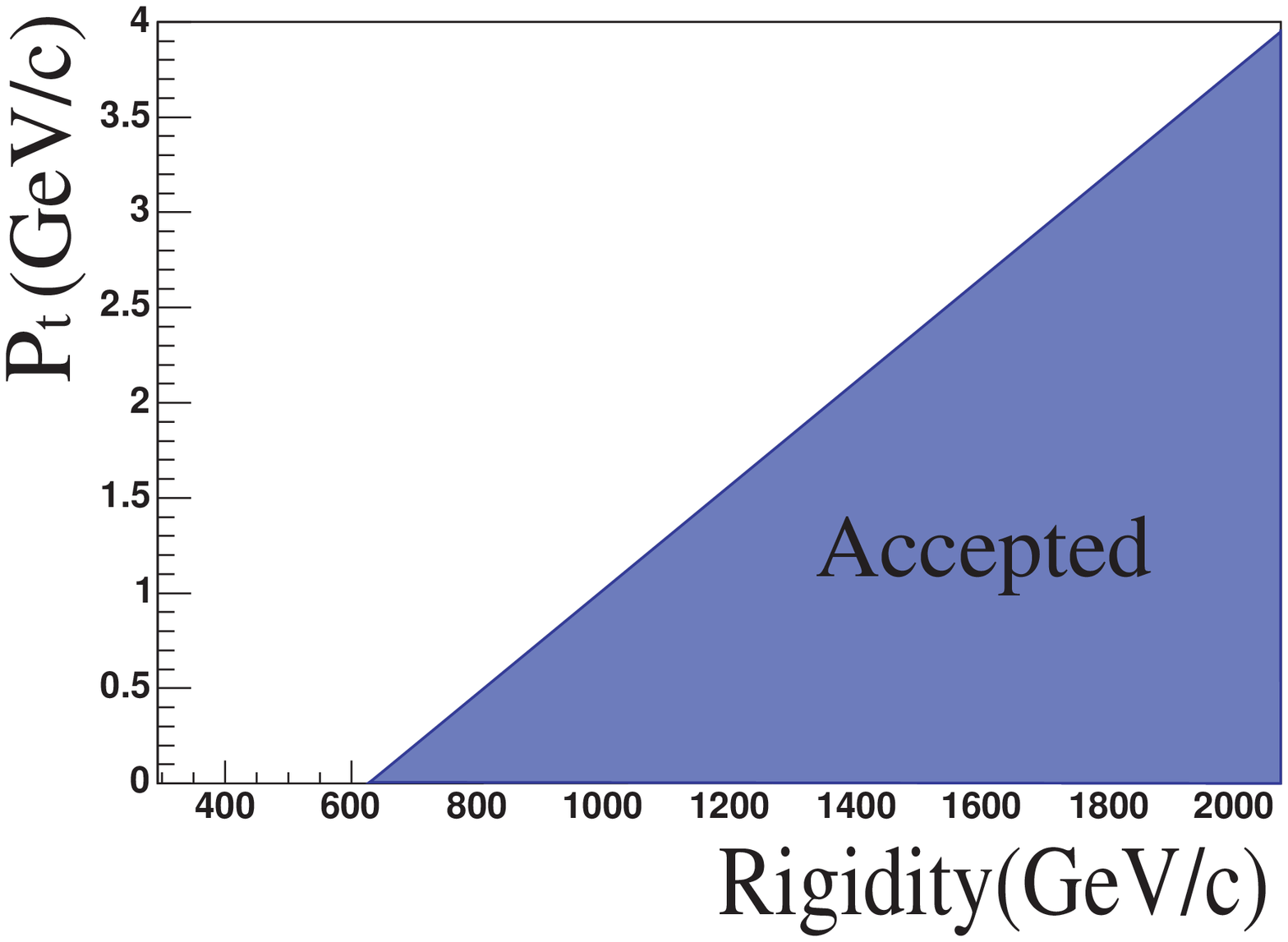}
\includegraphics{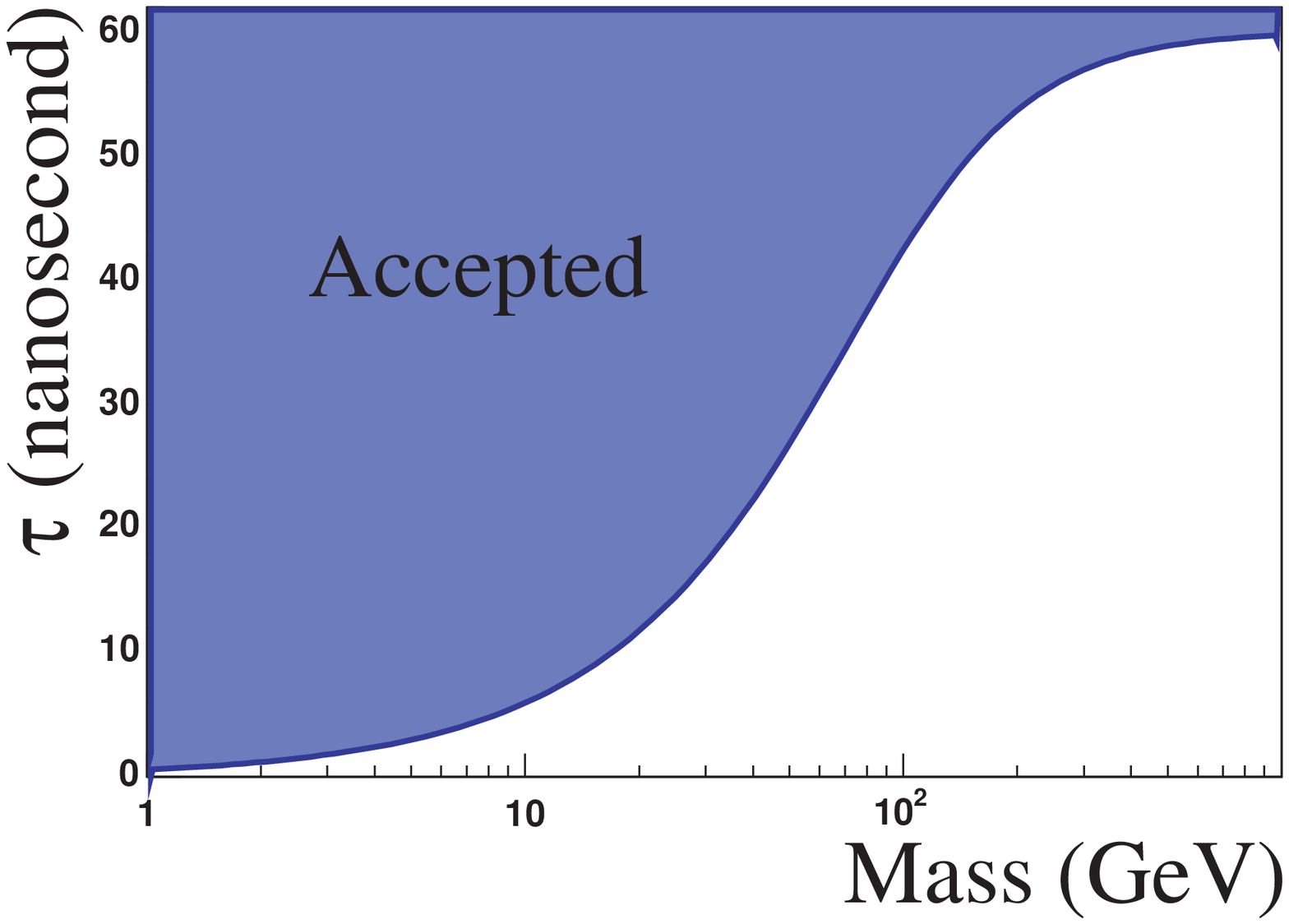}}
\caption{ (left) Acceptance at ZDCs particles with a azimuthal angle of 180 degrees in laboratory. (right) ZDCs' sensitivity to strangelets' proper life time, as a function of mass.\label{fig:acceptance}}
\end{center}
\vspace{-0.5cm}
\end{figure}

Since a strangelet consists of approximately equal numbers of $u$, $d$ and $s$ 
quarks, it can have a large baryon number and, 
a much lower charge-over-mass ratio than an
ordinary nucleus. At RHIC experiments, if a strangelet is produced in 
central Au+Au collisions where a QGP is expected to be created, it will
deposit a large signal in one of the Zero Degree Calorimeters(ZDCs)~\cite{ZDC}.
Particles  with small rigidity are swept away by strong magnetic field produced
by the dipole magnets, which are located in front of ZDCs and are used to bend 
beams. Only neutral or close-to-neutral particles, like strangelets, can 
survive the strong magnetic field and reach ZDCs. At ZDCs, the acceptance of 
strangelets depends on their rigidities and transverse momentum, and, 
because of the rectangular shape of ZDCs, slightly depends on their emitted
azimuthal angles. 
The left panel of Figure~\ref{fig:acceptance} shows an example of the 
acceptance. Although our set up requires a strangelet to have large rigidity 
in order to be detected, it requires loosely on strangelet's proper 
life time. The right panel of 
Figure~\ref{fig:acceptance} shows our sensitivity to strangelets' proper 
life time, as a function of mass.  While most efforts of strangelet 
search in heavy ion experiments look for strangelets with 
proper life time $> 50$ ns, we can detect strangelets that live much shorter 
than that. It is also worth to note that this is the first time to search
for strangelet at RHIC energies, which, in the center of mass frame, are at 
least four times larger than energies of strangelet search experiments done 
in the past. The other difference between our
efforts and past experiments is that, we focus at very forward production 
while most other experiments looked for strangelets around mid-rapidities.

\begin{figure}[ht]
\begin{center}
\resizebox{
\textwidth}{!}{
\includegraphics{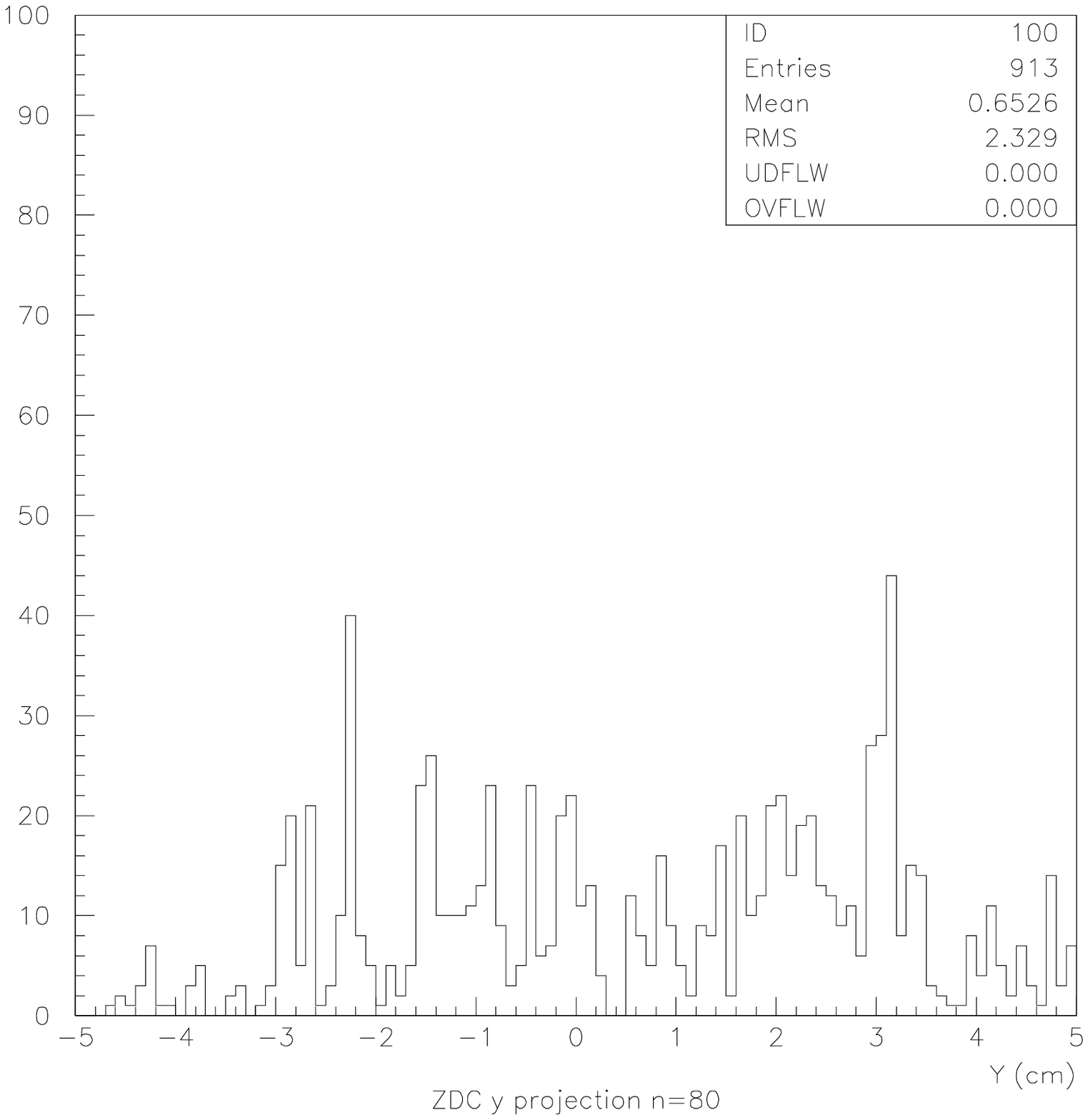}
\includegraphics{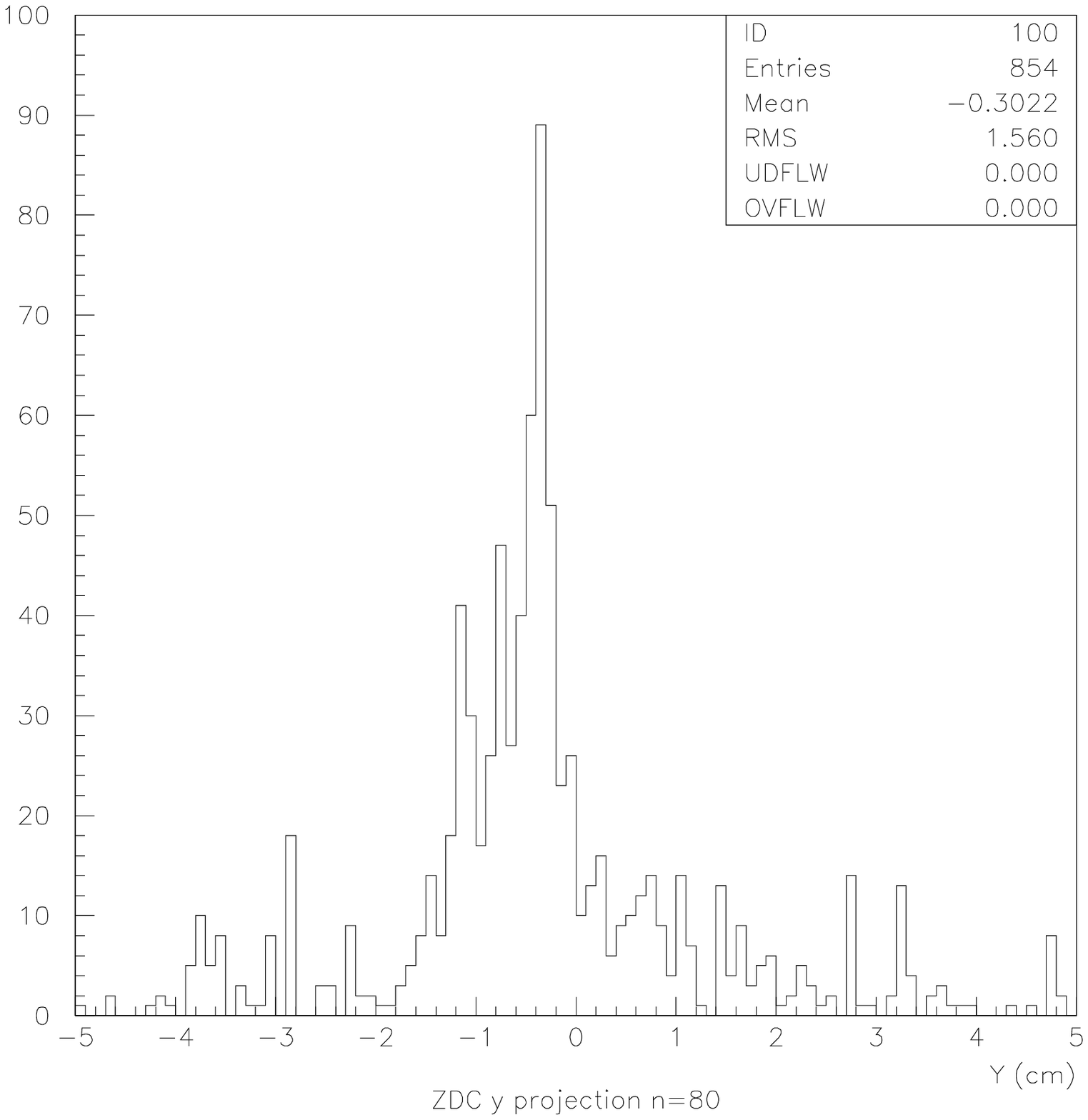}}
\caption{
Simulation of a normal event (left) and a strangelet event (right).
\label{fig:simulation}}
\end{center}
\vspace{-0.7cm}
\end{figure}

Note that a cluster of neutrons can give large signal in ZDC just like 
strangelets do, but the signals from neutron clusters
are more dispersed due to the fermi motion of spectator neutrons. 
This is shown by the Geant simulation in Figure~\ref{fig:simulation}. 
In the Figure, the hits deposition in one layer of 
the ZDC (out of 260 in total) is projected to the Y axis (both X and Y axes 
are perpendicular to the beam direction). Due to the normal $p_T$ 
distribution among spectator neutrons, the hits are dispersed along the 
Y axis. The same simulation repeated for a strangelet shows a prominent peak 
and less dispersion (right panel). Thus one can distinguish a strangelet 
event from normal events if, in addition to the total energy deposit in ZDCs,  
the transverse position information of energy deposition at ZDCs can be 
obtained.

In the Fall of 2003 STAR installed Shower Maximum Detectors (SMD)  
sandwiched (same acceptance as ZDCs) in between the first and second modules 
of each existing STAR ZDCs. Each SMD (east and west) consists of two 
scintillating plastic planes, one of 7 vertical slats of and another of 8 
horizontal slats. These two 
SMDs provide event-by-event information on the transverse position of the 
spectator neutrons produced in the collision thus allow us to perform the 
strangelet search. Besides strangelet search, the installation of ZDC-SMDs 
was motivated by a few other topics, which are beyond the interest 
of this paper, like flow, spin etc.

To maximize the possibility of of strangelet discovery, we select central 
events with high ZDC signals. As mentioned above, the strangelet
production through the distillation process via the creation of a QGP, 
which is more likely to happen in central events, is expected to have larger 
chance than the other two processes. Another reason of doing so is that 
ZDC signals for normal central events are less saturated, which makes 
it easy to identify an events with abnormally high ZDC signals.
During run 2004 two special triggers for strangelet search were implemented.
The trigger conditions at level zero (L0) are that, 
the signal from Central Trigger Barrel (CTB) has to be greater than 23000 
counts to select central events, and the sum of both ZDCs (east and west) 
signals has to be greater than 130 counts to select events that have large 
energy deposition in ZDCs.  
The total rejection obtained with this trigger is 99.97\% over minimum bias
events. In addition to the L0 trigger, a level three (L3) trigger was 
implemented to reject 70\% of the events that have passed the L0 trigger. The L3 
trigger is done by the following : for each ZDC, on the graph 
of ZDC signal versus total CTB sum, a curve is made above the band that 
consists of minimum bias events, with the curvature following the shape of 
the band. Events that appear to be below the curve are rejected. 
The curve can be shifted 
up and down to tighten or loosen the rejection. In total we have recorded 167k 
events with the L3 trigger, so far only 11.6k (7\% of the total) of them are 
reconstructed by the data production chain of STAR. For those events that are 
reconstructed, a Quality Assurance (QA) cut is applied to remove possible 
events from pile-up. The QA cut is made as the following : On the 
distribution of difference between the reconstructed 
position of primary z vertex (along the beam line) from offline tracking, 
and that from the timing information recorded by the Beam Beam Counters (BBCs),
events that appear to be two sigma (from the gaussian fit on the same 
distribution for normal events) away from zero are rejected.  This cut 
reduces the sample to four thousand.

\begin{figure}
  \begin{center}
\resizebox{!}{6.5cm}{
\includegraphics{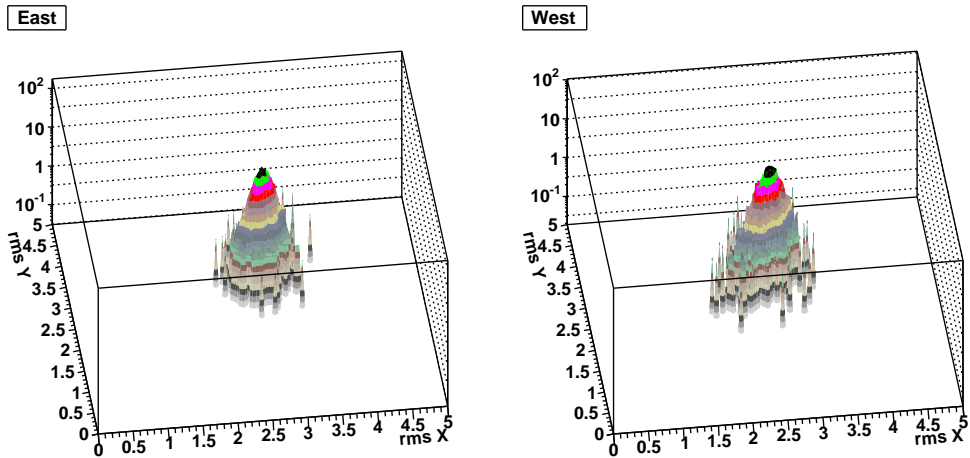}}
\caption{ The distribution of RMS from east SMD (left) and west SMD (right) \label{fig:rms}}
  \end{center}
\vspace{-0.7cm}
\end{figure}

The final result is presented in Figure~\ref{fig:rms}. The plot shows the distribution of
rms from SMD in Y direction versus that in X direction. If  
a strangelet is created in the collision and reaches one of the ZDC-SMDs, it
is expected to produce a spike in SMD and show up in the bottom-left corner 
in one of the two panels. However, the plot shows that the rms distribution
for both SMDs are well converged, no events with abnormally low rms are observed.
Based on that, collecting our events number and trigger rejections, 
we set a upper limit of 2x$10^{-8}$ with 90\% confidence level for the 
strangelet production in forward region at RHIC. Since the plot is based 
on only 7\% of the total statistics obtained, we might tighten that limit 
further in the future.

In summary, we have demonstrated the capability of STAR for strangelet search, 
with the recently installed ZDC-SMDs. Based on 7\% of the total statistics 
obtained during run 2004, we set up a upper limit for strangelet production of
2x$10^{-8}$ with 90\% confidence level. This is the first attempt at a 
strangelet search at RHIC.

{\it Acknowledgments---} Besides the acknowledgments in appendix, we wish to 
thank Sebastian White for consultations and help in building STAR ZDC-SMDs.



\end{document}